\documentclass[runningheads]{llncs}
\usepackage[T1]{fontenc}
\usepackage{amsmath}

\usepackage{url}
\usepackage{latexsym}
\usepackage{graphicx}
\usepackage{tikz}
\usepackage[font=scriptsize]{caption}

\definecolor{red}{RGB}{219, 0, 0}
\definecolor{grey}{RGB}{191, 191, 191}

\newcommand\T{\rule{0pt}{3ex}}       
\newcommand\B{\rule[-2ex]{0pt}{0pt}} 

\newcommand*\samethanks[1][\value{footnote}]{\footnotemark[#1]}

\begin{document}

\title{Learning Job Titles Similarity from Noisy Skill Labels}

\titlerunning{Learning Semantic Representations for Job Titles Using Noisy Skill Labels}

\author{Rabih Zbib \and
Lucas Alvarez Lacasa\and
Federico Retyk \and
Rus Poves\and
Juan Aizpuru \and \\
Hermenegildo Fabregat \and
Vaidotas Šimkus\thanks{Work done while at Avature.} \and
Emilia García-Casademont\samethanks[1]
}

\authorrunning{R. Zbib et al.}

\institute{Avature Machine Learning\\
\email{machinelearning@avature.net}\\}

\maketitle


\begin{abstract}

Measuring semantic similarity between job titles is an essential functionality for automatic job recommendations. This task is usually approached using supervised learning techniques, which requires training data in the form of equivalent job title pairs. In this paper, we instead propose an unsupervised representation learning method for training a job title similarity model using noisy skill labels. We show that it is highly effective for tasks such as text ranking and job normalization.


\end{abstract}


\section{Introduction}
\label{sec:intro}

With the significant growth of online platforms for job postings and applications, intelligent recommendation systems have become a necessity for both applicants and recruiters. Measuring semantic similarity between job titles is an essential functionality of these systems, whether to recommend suitable job openings to candidates, or vice versa. Job title similarity can be used as the sole measure for relevance, or more generally, as a component for computing an overall score between jobs and candidates, together with other information such as education, skills, and location. 
Recent work on this task goes beyond explicit lexical similarity and uses machine learning models that represent the semantics of the job title~\cite{mitra2017neural}. It takes the usual approach of learning semantic similarity by training a Siamese network using training data in the form of similar pairs~\cite{neculoiu2016}. But such labeled data is often hard to obtain in adequate amounts and quality. Skills, on the other hand, have recently become a major focus of the talent management industry~\cite{djumalieva12018}, as they are highly informative of candidate abilities and job requirements.

In this paper, we present {\em Job Similarity Training}, a new unsupervised technique for training a job title encoder for similarity tasks in two stages. We first compute a dense-vector \textit{auxiliary} embedding on sets of skills associated with a job title. Second, we train a neural job title encoder to mimic those auxiliary embeddings. The encoder maps a new job title into a dense vector that can be used for semantic similarity, text retrieval, and job normalization tasks. However, assigning accurate and comprehensive skills to candidate profiles or job descriptions is a specialized and time-consuming process, and automatic extraction of skills using a machine learning model also requires high-quality training data. We instead use {\em noisy} skills extracted from job listings and resumes with simple string matching. 

We conduct a series of experiments that compare our method with alternative approaches, and show that it is highly effective for retrieval and normalization. The contributions of this paper are summarized as follows:

\begin{itemize}
  \item {\em Job Similarity Training}, a new unsupervised training procedure for learning a job title encoder trained on noisy skills.
  
  \item A new evaluation dataset that we release as a benchmark for measuring the performance of job title encoders in a text ranking task.
  
  \item A set of experiments that show the effectiveness of the proposed unsupervised procedure for job title semantic similarity tasks, where we show that it outperforms a series of baselines, including the current state-of-the-art~\cite{JobBert2021}.
  
  \item Additional experiments, presented in the Appendices, which help to analyze the work from different perspectives: in Appendix \ref{sec:appendix-sample-efficiency} we study the efficiency of the training procedure, in Appendix \ref{sec:appendix-multilingual} we study the performance of our method in multilingual and cross-lingual text retrieval tasks, and in Appendix \ref{sec:appendix-predicting-skills} we study the extent to which the resulting job title encoder model captures information about the associated skills.
\end{itemize}


\section{Related Work}
\label{sec:rel-work}

Deep learning approaches for text ranking typically involve an encoder that converts the input text (query or document) into a dense, fixed-size representation, and then uses a similarity function to rank results. Huang et al.~\cite{huang2013} use an MLP-based architecture for the encoder, and cosine similarity for ranking. They train the MLP using a supervision signal from a downstream retrieval task. Palangi et al.~\cite{palangi2016} train an LSTM-based encoder to predict similarity of sentences, and apply it to text ranking within a web search engine, with user clicks as supervision. A similar approach, but using BERT~\cite{devlin-etal-2019-bert} with an attention layer, was explored by Humeau et al.~\cite{humeau2020}. For a detailed review of the use of BERT in information retrieval (IR) architectures, we refer the reader to Chapter 5 of the survey by Lin et al.~\cite{lin2021}. Our method is based on a similar scheme, except that we use an unsupervised training procedure instead of supervised learning.

There has recently been an increase in work relevant to tasks in the talent management domain. Javed et al.~\cite{carotene} normalize job titles using a hierarchy of successively more specific classifiers. Neculoiu et al.~\cite{neculoiu2016} present a supervised, end-to-end deep learning approach to the same problem. They train a character-based BiLSTM encoder to map job titles to a fixed-length vector, such that the distance between the vectors is a measure of the job title semantic similarity. The encoder is trained as a Siamese network with supervised contrastive loss. Zhang et al.~\cite{zhang2019} and Yamashita et al.~\cite{yamashita2022} present methods for learning job title representations by using graphs to model the relationship between jobs in candidate career paths, combining different ways of encoding the job title into a multi-view graph-based representation.

Finally, JobBERT by Decorte et al.~\cite{JobBert2021} is most similar to our approach. They use a pre-trained BERT language model to extract features for each token in the job title, and aggregate those features into a fixed-length vector using a soft attention layer. To train the job title encoder (including fine-tuning BERT) without a labeled dataset, they use the skills associated with each job title. Inspired by the Negative Sampling technique used for Skip-gram~\cite{mikolov2013}, they train the encoder as a classifier that predicts whether or not a particular skill corresponds to a job title. In our work, we use a different training objective and different neural architecture design. We compare the two approaches experimentally in Section~\ref{sec:experiments}.


\section{Proposed Method}
\label{sec:method}

\begin{figure*}[t!]
\includegraphics[width=1.0\textwidth]{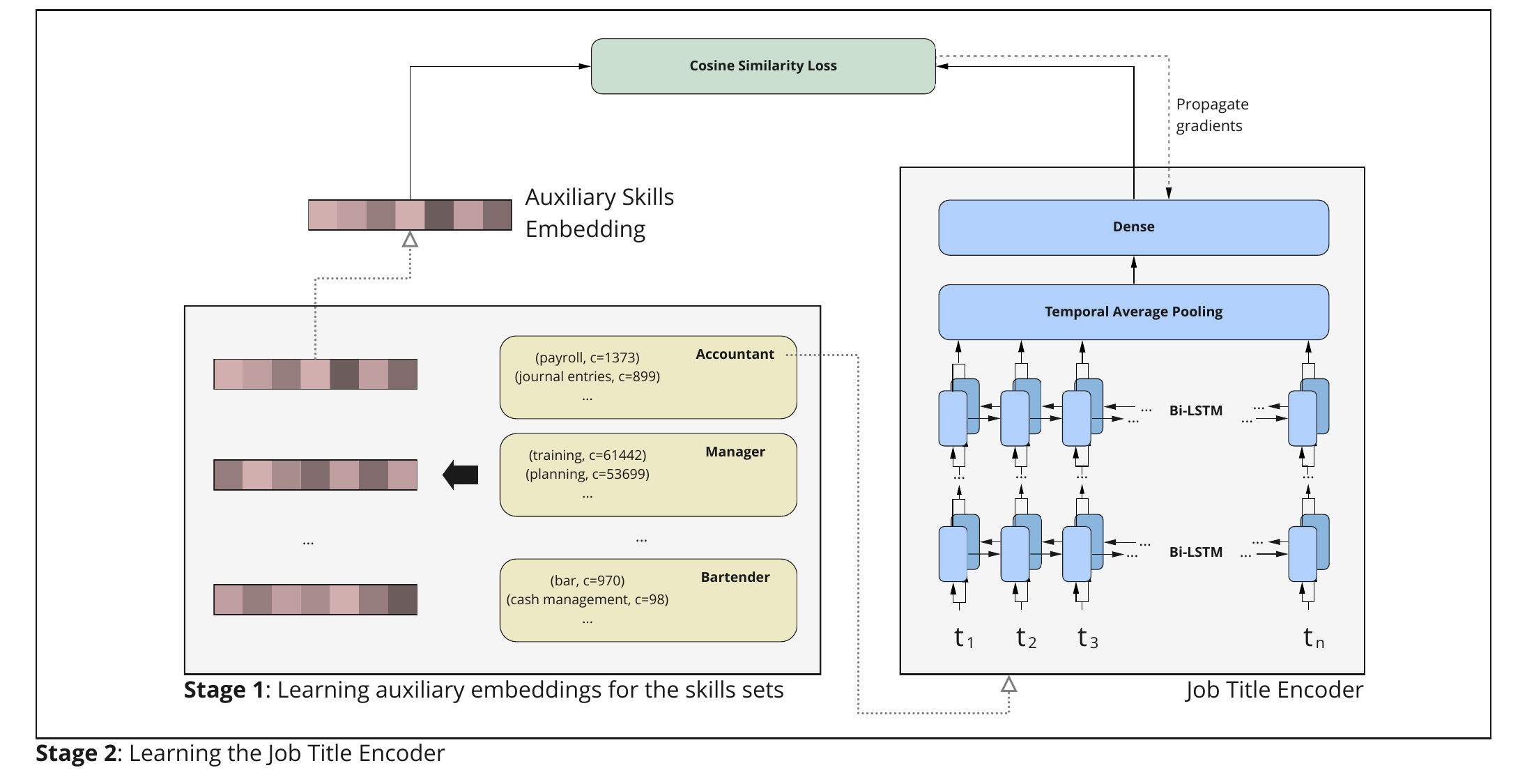}
\caption{Overview of {\em Job Similarity Training}, exemplified with a BiLSTM encoder. In {\bf Stage~1}, we obtain an auxiliary representation for each job title using the set of related noisy skills. These embeddings only use the skills for each job (i.e.~the job title text is not used so far). In {\bf Stage~2}, we train the job title encoder to map the job titles to their skills auxiliary representations. The trained encoder is used for downstream tasks, such as IR or job normalization.}
\centering
\label{fig:architecture}
\end{figure*}

We next present the paper's main contribution: \textbf{Job Similarity Training}, an unsupervised training procedure for learning semantic representations of job titles. We train a neural network encoder of job titles in two stages. We first learn an embeddings-based auxiliary representation for each job position using its noisy skills (and requiring no human-labeled supervision), and then train the encoder to map the job title text to its auxiliary representation (see details in Figure~\ref{fig:architecture}).




\subsection{Data Collection and Preprocessing}
\label{subsec:method-data}

The input to the entire training procedure is a dataset $\mathcal{D}_{\text{raw}}$ of pairs $(j, \mathbf{s}_{j})$, where $j$ is the job title text, $\mathbf{s}_{j}$ is a set of skills $s \in \mathcal{S}$, and $\mathcal{S}$ is a predefined skills vocabulary. In our experiments, the dataset is constructed from English job postings and from the work experience section of anonymized resumes. The skills are extracted from the description text using simple string matching, producing a noisy version of the skill set associated with each job.

Next, we generate the dataset $\mathcal{D}_{\text{merged}}$ by aggregating the noisy skill sets that correspond to the same job title. This aggregation results in a multiset $\mathbf{s}^{+}_{j} = \{s_1{:}\,\,m_1, \ldots, s_{n_j}{:}\,\,m_{n_j}\}$ where $s_i \in \mathcal{S}$ is a skill and $m_i$ is its count. Aggregating skill counts across common job titles reduces the noise since a skill that is extracted multiple times is more likely to be correct.

\subsection{Skills Auxiliary Representation}
\label{subsec:method-aux-repr}

In training Stage~1, we use the dataset $\mathcal{D}_{\text{merged}}$ to produce an auxiliary representation for each job title. We use the Doc2vec model~\cite{le2014} to directly compute a fixed-length vector $\textbf{e}_{j}$ for each skill multiset $\mathbf{s}^{+}_{j}$, considering it as a document and treating individual skills as words. The output is a dataset $\mathcal{D}_{\text{aux}}$ of pairs $(j, \textbf{e}_{j})$, with $j$ a job title  and  $\textbf{e}_{j}$ its auxiliary representation. These representations are then used as synthetic targets in the second training stage.

In the experiments, we use the Distributed Bag of Words (PV-DBOW) version of Doc2vec. PV-DBOW obtained better results than other approaches, including the Distributed Memory (PV-DM) version of Doc2vec as well as the average vector of individual skill embeddings learned with Word2vec~\cite{mikolov2013efficient}. Full details of the comparison are excluded for space considerations.

\subsection{Job Title Encoder}
\label{subsec:method-encoder}

In Stage~2 we train an encoder $\eta$ that maps the job title text to its final representation, motivated by the assumption that the auxiliary representations learned in the first stage capture important semantic information about the job titles $j$. Using the dataset $\mathcal{D}_{\text{aux}}$, the encoder is trained to minimize the cosine distance between $\eta(j)$ and the corresponding auxiliary representation $\textbf{e}_{j}$.

It is important to observe that, while the encoder $\eta$ is trained using information about the noisy skill sets (via the auxiliary representations $\textbf{e}_{j}$), the only information visible to the encoder during inference is the job title text. The encoder is thus forced to learn to infer information about the related skills. When provided with a new, unseen job title $j$, the model is able to generalize and infer about job-to-skills relationship. Similar to JobBERT~\cite{JobBert2021}, we also recognize that skills are useful information for building representations of job titles. But our unsupervised training procedure is markedly different: while they use a contrastive learning objective to train the encoder, we train it to mimic the representations $\textbf{e}_{j}$ obtained from the skills. We compare the two approaches in Section~\ref{sec:experiments}.


\section{Experiments}
\label{sec:experiments}

We next study the proposed approach through a series of experiments in two downstream tasks: text ranking and job normalization. As described in Section~\ref{sec:method}, for the training data we extract job titles and skills from job postings and anonymized resumes.  The resulting raw dataset $\mathcal{D}_{\text{raw}}$ contains 44 million samples, and the merged set $\mathcal{D}_{\text{merged}}$ 8.5 million sample pairs. The set of skills we use contains 5,600 skills covering a variety of industries.

\subsection{Text Ranking}
\label{subsec:experiments-ranking}

Text ranking involves ranking a set of text documents in a corpus given a query and a measure of relevance. In the context of a job title recommendation system, both the documents and the queries are short text. The corpus job titles are ranked by first computing a vector representation using the encoder $\eta(j)$, and then computing the cosine distance between the query and the job title vectors.

We use a new job title similarity dataset\footnote{\url{https://github.com/rabihzbib/jobtitlesimilarity_dataset}} for evaluation, consisting of 2,724 job titles from different industries (105 of which are used as queries). Each query/corpus-item pair is labeled for binary relevance after adjudicating two independent human annotations\footnote{With an inter-annotator agreement measured at 86\%}. The main experimental results are in Table~\ref{tbl:english-results}. Following common practice, we evaluate models using the Mean Average Precision (MAP) of the output of the ranked lists, as well as Precision at 5 and 20. We use the \texttt{trec\_eval} software library to compute these metrics\footnote{\url{https://github.com/usnistgov/trec_eval}}.

\begin{table}[t]
\scriptsize
\begin{center}
\begin{tabular}{|ll|ccc|}
\hline 
\multicolumn{2}{|l|}{\bf \T Method} & \bf MAP & \bf P@5 & \bf P@20\\

\hline
\multicolumn{2}{|l|}{\T \textbf{Text-based Retrieval}} & & & \\
\bf Model & \bf Training Method & & & \\
~~Okapi BM25          & ~~Trained on $\mathcal{D}_{\text{merged}}$  & 0.2754 & 0.5067 & 0.3062      \\ 

~~BERT             \B & ~~None (no fine-tuning)             & 0.1556  & 0.3124  & 0.1871  \\ 

~~BiLSTM             & ~~Negative Sampling     & 0.6428  & 0.7581  & 0.5376    \\ 
~~BERT            \B & ~~Negative Sampling     & 0.6011  & 0.7238  & 0.5152   \\ 

~~BiLSTM             & ~~Job Similarity Training       & 0.6814  & 0.7790  & 0.5781   \\ 
~~BERT            \B & ~~Job Similarity Training      & \bf 0.7077  & \bf 0.7829  & \bf 0.5929    \\ 

\hline
\multicolumn{2}{|l|}{\T \textbf{Skill-based Retrieval}} & & & \\
\multicolumn{2}{|l|}{~~TF-IDF (Noisy Test Skills)}             & 0.3319      & 0.5481       & 0.3135       \\ 
\multicolumn{2}{|l|}{~~Doc2vec (Noisy Test Skills)}         \B & 0.1031      & 0.1675       & 0.1204      \\ 

\multicolumn{2}{|l|}{~~TF-IDF (Gold Standard Test Skills)}            & 0.7880  & 0.8376  & 0.6668       \\ 
\multicolumn{2}{|l|}{~~Doc2vec (Gold Standard Test Skills)}        \B & 0.7126  & 0.7446  & 0.5921       \\ 

\hline
\end{tabular}
\end{center}
\caption{Performance of different unsupervised models on the text ranking evaluation set. Mean Average Precision (MAP), Precision at 5, and Precision at 20 are shown for each condition. Bold font style corresponds to the method achieving the best results.}
\label{tbl:english-results}
\end{table}


\noindent{\textbf{Text-based Retrieval}}.The {\em Job Similarity Training} experiment use two different architectures for the encoder $\eta$: a BiLSTM encoder trained from scratch, with SentencePiece~\cite{DBLP:conf/emnlp/KudoR18} for tokenization and average pooling to aggregate the token representations; and a pre-trained BERT-base encoder~\cite{devlin-etal-2019-bert} using its built-in pooling mechanism.  Both architectures have a linear Dense layer on top to produce a vector with the same dimensions as the auxiliary skills representations.

We compare {\em Job Similarity Training} to two simple baselines. The first is the well-known ranking function Okapi BM25~\cite{jones2000a,INR-019}, trained and evaluated using the job title text and ignoring the skills. The second baseline is a pre-trained BERT-based language model with no fine-tuning. Table~\ref{tbl:english-results} shows that both baselines obtain MAP scores well below the models trained with either Job Similarity Training or Negative Sampling. These results support the hypothesis that using skill-related information is important to train the encoder effectively.

{\em Job Similarity Training} is also compared to a similar unsupervised alternative: the Negative Sampling scheme proposed by Decorte et al.~\cite{JobBert2021} to train JobBERT. In this case, the encoder is trained to predict whether a skill is related to a job title. Results show that our method outperforms Negative Sampling for both architectures. For a fair comparison, the same two encoder architectures are used for both methods.


\noindent{\textbf{Skill-based Retrieval}}. A natural question in the context of this work is how effective would it be to rank jobs by looking only at their set of skills, and not the job title text. To answer this, we use the skills of the test set jobs to build two vector representations: discrete, high-dimensional TF-IDF skill vectors, and Doc2vec embeddings of the skill sets (i.e., the synthetic target $\textbf{e}_{j}$ used to train $\eta(j)$). In both cases, the skill vectors are used directly to compute the cosine distance\footnote{In this paper's method, as in JobBERT~\cite{JobBert2021}, skills are not needed at inference time. The test job skills are used here for comparison only.}. We use two versions of the test job skills:

\begin{itemize}
    \item \textit{Test Job Noisy Skills}. For each job title in the test set, we extract a skill set from its description text, consistent with how noisy skills are extracted for training.
    \item \textit{Gold Standard Skills}. For each job title in the test set, Talent Management experts define a skill set. This version is intended to measure the effect of the quality of the skills on retrieval results.
\end{itemize}

We report on two experiments for each of these two sets, using Doc2vec and TF-IDF vectors. The performance of the Noisy Skill-based baselines is significantly lower than that of our method, showing that using the noisy skills directly to represent the job at inference time is not adequate. The Gold Standard baselines outperform our methods, but they use the gold-standard skill set to represent each test jobs. From a practical perspective, the difference between the Noisy and the Gold Standard baselines shows that high-quality skills can be directly used for retrieval, but they are often not available in practice. With noisy skills, however, the performance is significantly degraded. Our method leverages the noisy skills to learn a semantic representation of the job titles. The model \emph{generalizes} to unobserved job titles, as shown by the significantly better performance that the job title encoder achieves compared to the Noisy Skills baselines.

\subsection{Job Normalization}
\label{subsec:experiments-normalization}
The second downstream task is job normalization, which consists of mapping an input job title to one element in a set of normalized job titles. Here, we also encode the raw input job as well as the normalized job titles, and select the one corresponding to the smallest cosine distance. We use the test set of Decorte et al.~\cite{JobBert2021}, which consists of 15,463 raw job titles (as typically found on online job boards) and their corresponding normalized job title. 2,675 ESCO occupations\footnote{\url{https://esco.ec.europa.eu/en/classification/occupation_main}} are used as the normalized set. Table~\ref{tbl:decorte-results} shows that the Job Similarity Training models also outperform the results reported by JobBERT for job normalization.

\begin{table}[t]
\scriptsize
\begin{center}
\begin{tabular}{|ll|ccc|}
\hline 
\bf \B\T Model & \bf Training Method & \bf MRR & \bf P@5 & \bf P@10 \\
\hline
 BERT          \B\T & Decorte et al.~\cite{JobBert2021}  & 0.3092  & 0.3865  & 0.4604 \\  
 BiLSTM             & Job Similarity Training                & 0.3007  & 0.3955  & 0.4760 \\ 
 BERT            \B & Job Similarity Training                & \bf 0.3414  & 0.4595  & 0.5400 \\ 
\hline
\end{tabular}
\end{center}
\caption{Performance of different unsupervised models on Decorte~\cite{JobBert2021}'s test set. Mean Reciprocal Rank (MRR) and Precision at 5 and 10 of the micro averages are shown.}
\label{tbl:decorte-results}
\end{table}


\section{Conclusion}
\label{sec:conclusion}

In this paper, we presented {\em Job Similarity Training}, a new unsupervised technique for training job title encoders to learn semantic information. We evaluated it on two downstream tasks: text ranking, using a new evaluation dataset that we release as part of this work, and job normalization, using an existing benchmark.

Our experiments show that the new procedure outperforms various baselines in both tasks, including JobBERT (the previous state-of-the-art). Furthermore, they show that, while the quality of the skill-to-job mapping is a critical factor in the performance of systems that directly use skills for understanding the semantics of a job title, our method can leverage noisy skills to obtain good results.

We also presented additional experiments (included in the Appendices) where we explore the efficiency of the method when compared to JobBERT, its performance in multilingual and cross-lingual scenarios, and the encoder's ability to capture information about skills related to a job title.

\newpage

\bibliographystyle{splncs04}
\bibliography{bibliography.bib}


\appendix

\newpage

\section{Analysis of Training Procedure Efficiency}
\label{sec:appendix-sample-efficiency}

\begin{figure}
\includegraphics[width=1.0\textwidth]{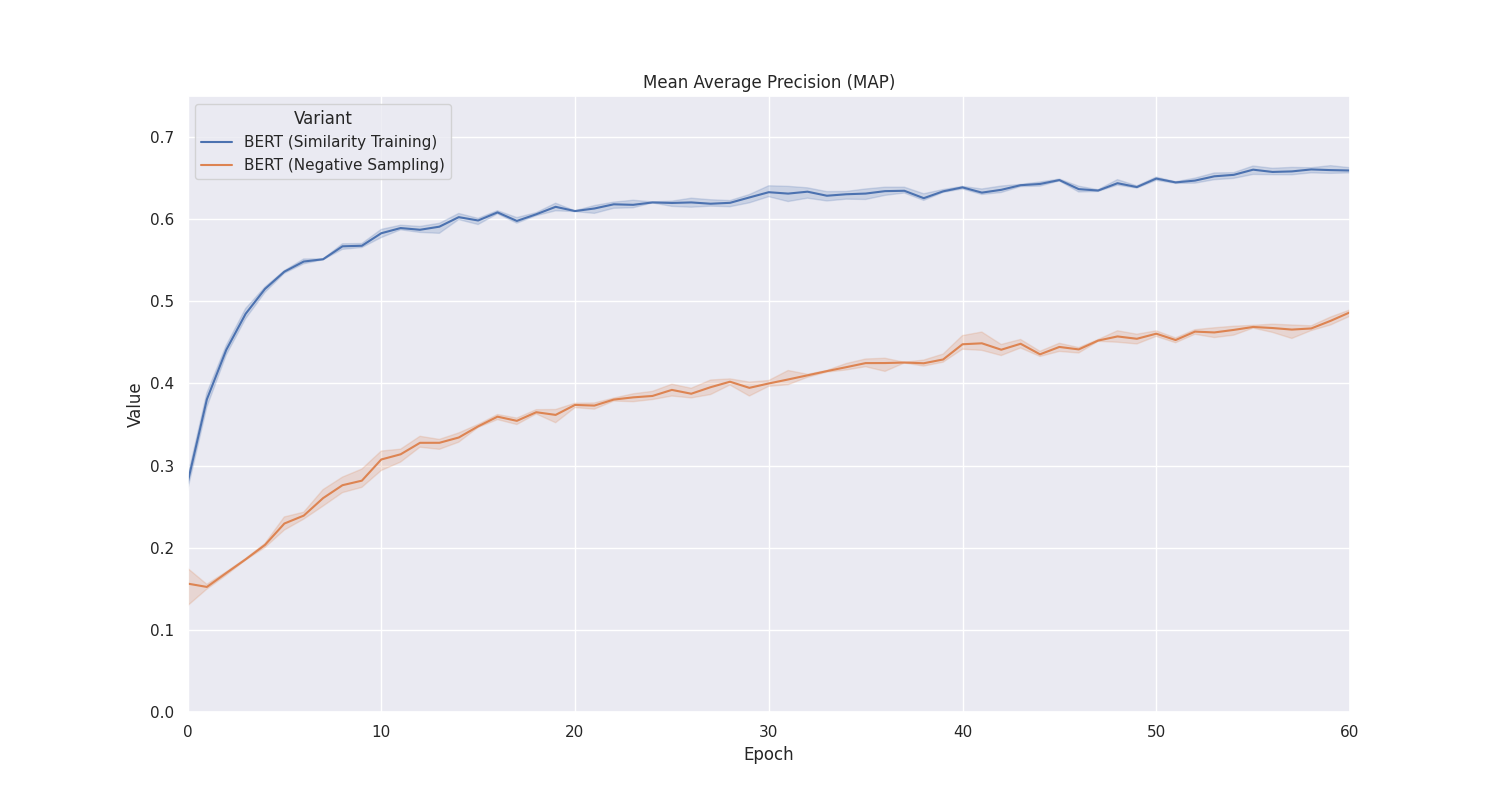}
\caption{Performance, as measured in a downstream text ranking task, for different model snapshots during the unsupervised training procedures: Job Similarity Training and Negative Sampling.}
\centering
\label{fig:efficiency}
\end{figure}

In this section, we analyze the difference between Job Similarity Training and Negative Sampling in terms of efficiency. In order to do this, we fine-tune a BERT-small encoder with each unsupervised method (Similarity Training, Negative Sampling), and periodically measure the performance of these models in the text ranking task. The encoder architecture, the training hyperparameters, and the number of steps per epoch are identical in both cases. The encoder trained with Job Similarity Training uses $\mathcal{D}_{\text{merged}}$ as its training dataset. The one trained with Negative Sampling uses positive samples from the dataset $\mathcal{D}_{\text{raw}}$ and generate negative samples according to that method's definition. Figure~\ref{fig:efficiency} compares the performance of both methods under these conditions. We aggregate the learning curves of three replications for each method.

Besides obtaining a superior performance at convergence, Job Similarity Training achieves a better MAP score much earlier in the training process. In other words, given a fixed amount of forward passes, our method performs significantly better in the downstream task. This is particularly important when fine-tuning large Transformer-based language models like BERT.

\section{Multi- and Cross-lingual Text Retrieval}
\label{sec:appendix-multilingual}

Although in our experiments, skills are extracted from English job descriptions, those skills represent language-independent concepts. This implies that the auxiliary representations that we use to train the job title encoder $\eta$ are also language-independent. So all that is needed to build an encoder in a new language is to translate the job titles $j$ included in each sample from the dataset $\mathcal{D}_{\text{aux}}$ to obtain a new dataset $\mathcal{D}^{\textbf{L}}_{\text{aux}}$ for the target language $\textbf{L}$. The corresponding model $\eta_{\textbf{L}}$ can thus be created without manual data effort and without repeating the initial data collection and preprocessing workload, which might be harder in the case of low-resource languages. Also, the language independence of the semantic space can be further exploited to build multilingual models by training the job title encoder $\eta$ on the combined data from multiple languages. The resulting model can be used for cross-lingual retrieval.

In this experiment, we take the original English dataset $\mathcal{D}_{\text{aux}}$ and use Machine Translation (MT) to translate around 1 million job titles $j$ to German and 2 million job titles $j$ to French, in order to obtain $\mathcal{D}^{\textbf{DE}}_{\text{aux}}$ and $\mathcal{D}^{\textbf{FR}}_{\text{aux}}$ respectively. These are used to train either monolingual or bilingual models $\eta$ with the same BiLSTM architecture explored in Section~\ref{subsec:experiments-ranking}, and trained with the proposed Similarity Training procedure.

To evaluate these models, we translate the evaluation set using human translators. The rationale is that, using MT for the training set, one can transfer the model to another language quickly, even if original data in that language is not available. On the other hand, producing high-quality human translations of the evaluation set is a better reflection of the real scenario, where users input native job titles rather than machine-translated ones.

The results in Table~\ref{tbl:french-german-results} show that this approach produces a viable model at no extra human effort in translation or labeling. The results for the German and the French monolingual models (MAP 0.5545 and 0.5355, respectively) are much higher than the Noisy-skills baselines from Table~\ref{tbl:english-results}. The encoder-based English models have better performance, as they were trained on 8.5 million samples, while the German models were trained on 1 million and the French models on 2 million.

We also experiment with bilingual models, combining data from two languages to train a French-English encoder and a German-English encoder. We observe that the bilingual model can be used for cross-lingual retrieval ---when either the query or the corpus titles are in English--- without loss in performance, providing evidence for the assumption that the information that can be captured from the skill set is correlated with the underlying, language-independent job position concept, and therefore is transferable across languages. The cross-lingual retrieval capability is achieved with no penalty in monolingual retrieval quality, since we see a small difference of less than 1\% between monolingual and bilingual models for retrieval when the query and the corpus are in the same language (first and second lines in each section of the table).  In practice, adding English to models in another language is a useful capability, since job titles in general are likely to include English terms.

\begin{table}
\begin{center}
\small
\begin{tabular}{|l|l|l|ccc|}
\hline \bf Training & \bf Q\textsubscript{Test} &\bf C\textsubscript{Test} &\bf MAP & \bf P@5 & \bf P@20 \\
\hline
\bf DE     & \bf DE & \bf DE & 0.5545 & 0.7000 & 0.4808 \\  
\bf DE+EN  & \bf DE & \bf DE & 0.5476 & 0.6885 & 0.4745 \\  
\bf DE+EN  & \bf DE & \bf EN & 0.5942 & 0.6712 & 0.5202 \\  
\bf DE+EN &  \bf EN & \bf DE & 0.5545 & 0.7010 & 0.4890 \\  
\hline
\bf FR     & \bf FR & \bf FR & 0.5355 & 0.6692 & 0.4750 \\  
\bf FR+EN  & \bf FR & \bf FR & 0.5437 & 0.6923 & 0.4707 \\  
\bf FR+EN  & \bf FR & \bf EN & 0.5908 & 0.6827 & 0.4933 \\  
\bf FR+EN &  \bf EN & \bf FR & 0.5577 & 0.6857 & 0.4867 \\  
\hline
\end{tabular}
\end{center}
\caption{\small Monolingual and bilingual model results on either monolingual or cross-lingual text retrieval tasks. For all experiments, we use the {\em Doc2vec (Noisy Skills) + Encoder} variant of the training procedure. Language codes: German (DE), English (EN), and French (FR).}
\label{tbl:french-german-results}
\end{table}

\newpage

\section{Predicting Skills from the Encoded Job Title}
\label{sec:appendix-predicting-skills}

Lastly, we discuss the results of a different kind of experiment, where the encoded job titles are used to predict related skills. The aim is to further show that the encoded job title representation captures information about the skills. This functionality is also useful to suggest possible skills for a user when creating a job description.

A consequence of the proposed training procedure is that the representation spaces of the encoded job titles and the auxiliary embeddings are compatible, in the sense that measuring the distance between an encoded job $\eta(j)$ and the skill auxiliary embeddings is meaningful. The same argument applies for the individual skill embeddings produced by Skip-gram and word2vec.

In this experiment, we use the job title encoder $\eta$ to encode the job titles from the evaluation set, and for each job title we retrieve the $N$ skills whose individual skills embeddings are closest in the shared representation space. Finally, we measure the precision of the predicted top-$N$ skills using the gold standard set of skills for that job title (the gold standard used for the skill-based baselines in Section~\ref{subsec:experiments-ranking}). We obtain the following results: P@5 is 0.4577, P@10 is 0.3854, and P@20 is 0.3127. Considering that generating a set of elements from a large vocabulary ---in this case, 5,600 skills--- is a difficult problem, even for humans, these results show that the encoder $\eta$ captures information about the skills effectively.

Furthermore, to allow for a qualitative assessment of the extent to which the model generalizes to new job titles, we show the predicted skills for two job titles that are not included in the training set. 

{
\small
\begin{verbatim}
Director of Communications:
    Public Relations, Social Media, Media Relations, 
    Campaigns, Strategy, Writing, Editing

Construction and Building Inspector:
    Construction, Inspection, Materials,
    Contractors, Structures, Plumbing,
    Concrete, Regulations, Completion, Welding
\end{verbatim}
} 

The second example is particularly interesting since the position involves the intersection of two concepts (\textit{Construction/Building}  and \textit{Inspector}), and the predicted skills reflect that.

\end{document}